\documentclass[reprint,twocolumn,aps,prb,showpacs]{revtex4-1}

\usepackage{amsmath}
\usepackage{amssymb}
\usepackage{graphicx}
\usepackage{dcolumn}
\usepackage{bm}
\usepackage[colorlinks=true, allcolors=blue]{hyperref}
\usepackage{epstopdf}
\usepackage[utf8]{inputenc}
\usepackage{amssymb}
\usepackage[dvipsnames]{xcolor}

\begin{document}
\preprint{APS/123-QED}

\title{Quantitative Determination of the Critical Points of Mott Metal-Insulator Transition
in Strongly Correlated Systems}


\author{Yuekun Niu$^{1}$}
\email[]{ykniu@imu.edu.cn}%
\author{Yu Ni$^{2}$}
\author{Jianli Wang$^{3}$}
\author{Leiming Chen$^{3}$}
\author{Ye Xing$^{3}$}
\author{Yun Song$^{4}$}
\email[]{yunsong@bnu.edu.cn}%
\author{Shiping Feng$^{4}$}
\email[]{spfeng@bnu.edu.cn}%

\affiliation{$^{1}$School of Physical Science and Technology, Inner Mongolia University, Hohhot 010021, China}

\affiliation{$^{2}$College of Physics and Electronic Information, Yunnan Normal University, Kunming 650500, China}

\affiliation{$^{3}$School of Materials Science and Physics, China University of Mining and
Technology, Xuzhou 221116, China}

\affiliation{$^{4}$Department of Physics, Beijing Normal University, Beijing 100875, China}

\begin{abstract}
Mottness is at the heart of the essential physics in a strongly correlated system as many
novel quantum phenomena occur in the metallic phase near the Mott metal-insulator transition.
We investigate the Mott transition in a Hubbard model by using the dynamical mean-field theory
and introduce the local quantum state fidelity to depict the Mott metal-insulator transition.
The local quantum state fidelity  provides a convenient approach for determining
the critical point of the Mott transition. Additionally, it presents a consistent description
of the two distinct forms of the Mott transition points.
\end{abstract}

\maketitle

\section{Introduction}
The Mott metal-insulator transition (MIT) \cite{Mott1949,Mott1968,Imada1998},
resulting from the interply between the kinetic energy $t$ and the on-site Coulomb repulsive interaction $U$
among electrons, represents a fundamental manifestation of strong electron correlation effects.
Experimental investigations have demonstrated the presence of the unconventional superconductivity and other
exotic quantum phenomena in the metallic phase  close to the Mott MIT \cite{Imada1998}.
This is why the quantitative determination of the critical point of the Mott
MIT is crucial to deeply understanding the essential physics of these novel quantum phenomena in the strongly
correlated systems.

Although enormous efforts have been made at the experimental and theoretical levels to understand
the physical origin of the Mott MIT, together with the associated novel quantum phenomena
\cite{Imada1998}, the quantitative determination of the critical point of the Mott MIT is still a
challenging issue. In early studies, it was shown in the Gutzwiller approximation that the
quasiparticle coherent weight can be used as a physical quantity to determine the critical point
of the Mott MIT, where the quasiparticle coherent weight $Z_{\rm F}$ disappears and the effective mass
diverges as $1/Z_{\rm F}$ when the strength of the Coulomb interaction approaches its critical value\cite{Brinkman1970,Bulla1999,Bulla2000}.
The quasiparticle coherent peak around the Fermi surface comes mainly from the scattering of electrons on the local-spin fluctuations. Hence, its disappearance at the critical point of MIT can also be tracked
by analyzing the energy dependence of the electron self-energy with different Coulomb repulsive
interactions\cite{Turkowski2021}. Later, a systematic analysis\cite{Georges1996} based on the
dynamical mean-field theory (DMFT) indicated that at low-temperature, the opening of the gap and the
vanishing of the quasiparticle coherent peak do not happen at the same critical value of $U_{\rm c}$.
Instead, MIT is found as a function of $U/t$, with the corresponding metallic and insulating solutions
coexisting between $U_{\rm c2}$ and $U_{\rm c1}$, respectively. Since then, a series of studies
focusing on the region of the metallic and insulating solutions coexisting between $U_{\rm c2}$ and
$U_{\rm c1}$ has been made
\cite{Florens2002,Feldbacher2004,Raas2009,Sordi2011,Eisenlohr2019,Zhou2020,Loon2020}.
In practice, these studies also indicate that the quasiparticle coherent weight $Z_{\rm F}$  may
not be able to mark out these two distinct forms of the critical points in the MIT
due to the coexistence of a branch of metastable metallic solution that connects the two
stable metallic and insulating solutions of $Z_{\rm F}$ \cite{Chatzieleftheriou2022, Ono2003}.
In this case, a natural question is raised: is there a more proper physical quantity to present the
existence of the two distinct forms of the critical points in the Mott MIT?

In this paper, we study the one-band Hubbard model by using the DMFT with the Lanczos method as its impurity solver.
It is confirmed that the local quantum state fidelity (LQSF),
as a proper physical quantity, can provide a convenient way to identify the critical point of the Mott MIT.
In particular, it can give a consistent description of the two different forms of the critical points in the Mott MIT.
	
\section{Models and Methods}
The one-band Hubbard model is the simplest model that captures the essential physics of MIT in a strongly
correlated system. The Hamiltonian of the one-band Hubbard model is given by
\cite{Kanamori1963,Hubbard1963,Hubbard1964,Hubbard1964j}
\begin{eqnarray}\label{hubbard-model}
\label{hubbard}
H=-t\sum_{\langle ij \rangle \sigma}d_{i\sigma}^{\dag}d_{j\sigma}
-\mu\sum_{i\sigma}d_{i\sigma}^{\dag}d_{i\sigma}
+\frac{U}{2}\sum_{i\sigma}n_{i\sigma}n_{i\bar{\sigma}},
\end{eqnarray}
where the summation $\langle ij \rangle$ is over all sites $i$, and for each site $i$, restricted
to its nearest-neighbor (NN) sites $j$, $t$ denotes the electron NN hopping amplitude, $U$ is the
on-site Coulomb repulsion between electrons, and $\mu$ is the chemical potential.
$d_{i\sigma}^{\dag}$ ($d_{i\sigma}$) is the creation (annihilation) operator for an electron with
spin $\sigma$ at lattice site $i$, and $n_{i\sigma}$ is the occupation number operator of
electrons at lattice site $i$. Unless explicitly stated, we set $t=1$ as the energy scale in this
paper. The electron Green's function of the Hubbard model (\ref{hubbard-model}) can be expressed
formally as
\begin{eqnarray}\label{lg}
\mathcal{G}_{\sigma}(\textbf{k},\omega)=\frac{1}{\omega+\mu-\varepsilon_{\textbf{k}}
-\Sigma_{\sigma}(\textbf{k},\omega)},
\end{eqnarray}
where the energy dispersion in the tight-binding approximation can be obtained directly
by $\varepsilon_{\textbf{k}}=\sum_{ij}t_{ij}e^{i\textbf{k}\cdot(\textbf{R}_i-\textbf{R}_j)}$,
while the effect of interaction
in the Hubbard model (\ref{hubbard-model}) has been encoded in the electron self-energy
$\Sigma_{\sigma}(\textbf{k},\omega)$. It should be emphasized that in the infinite dimensional
system, this electron self-energy $\Sigma_{\sigma}(\textbf{k},\omega)$ is momentum independent.
The DMFT \cite{Muller1989,Metzner1989} provides an approximate solution to this electron
self-energy $\Sigma_{\sigma}(\textbf{k},\omega)$ in a finite dimensional system by setting
$\Sigma_{\sigma}(\textbf{k},\omega)=\Sigma^{\rm (AIM)}_{\sigma}(\omega)$, where the momentum
independence of the electron self-energy
$\Sigma^{\rm (AIM)}_{\sigma}(\omega)$ can be obtained in terms of an auxiliary impurity model consisting of a single interacting site in a self-consistently determined bath\cite{Loon2020}.
In other words, the auxiliary impurity model provides a way to calculate the local electron
self-energy $\Sigma^{\rm (AIM)}_{\sigma}(\omega)$ and to use the entire repertoire of the
electron Green's function with the contribution to the electron self-energy taken from the
auxiliary impurity system rather than from a perturbation expansion\cite{Richard2016}.

We evaluate the electron self-energy of the Hubbard model (\ref{hubbard-model})
by using the DMFT with the {\it Lanczos} method as its impurity solver. In the framework of DMFT,
the Hubbard model (\ref{hubbard-model}) is mapped onto an effective single impurity model by
dropping the nonlocal contribution to the electron self-energy,
\begin{eqnarray}\label{anderson}
H_{imp}&=&\sum_{m\sigma}\varepsilon_{m}c_{m\sigma}^\dag c_{m\sigma}
+\sum_{m\sigma}V_{m}(c_{m\sigma}^\dag d_{\sigma}+d_{\sigma}^\dag c_{m\sigma})\nonumber\\
&+&\sum_{\sigma}(\varepsilon-\mu)d_{\sigma}^\dag d_{\sigma}
+\frac{U}{2}\sum_{\sigma}n_{d\sigma}n_{d\bar{\sigma}},
\end{eqnarray}
which becomes exact in the limit of the infinite lattice coordination \cite{Georges1992}.
Here $d_{\sigma}^{\dag}$ ($d_{\sigma}$) creates (annihilates) a particle in the impurity
orbital and $c_{m\sigma}^{\dag}$ ($c_{m\sigma}$) creates (annihilates) an electron in a
conduction band, where the impurity orbital and conduction band are coupled
each other via effective parameters $\varepsilon_{m}$ and $V_{m}$, which are determined by the
self-consistent DMFT calculation utilizing an impurity solver. In the following discussions, we
introduce the local electron Green's function in real-space as \cite{Anisimov2010,Mahan2000}
\begin{equation}\label{local-electron-Green-function}
\mathcal{G}_{\sigma}(\tau)=-<T_{\tau}d_{\sigma}(\tau)d_{\sigma}^{\dag}(0)>,
\end{equation}
with the imaginary time $\tau=it$. This local electron Green's function
(\ref{local-electron-Green-function}) in energy space can be obtained directly by performing the
Fourier transformation
\begin{eqnarray}
\mathcal{G}_{\sigma}(i\omega_{n})&=&\int_{0}^{\beta}d\tau e^{i\omega_{n}\tau}
\mathcal{G}_{\sigma}(\tau),\\
\mathcal{G}_{\sigma}(\tau)&=&\frac{1}{\beta}\sum_{n=-\infty}^{\infty}e^{-i\omega_{n}\tau}
\mathcal{G}_{\sigma}(i\omega_{n}),\label{gg}
\end{eqnarray}
where $-\beta\leq\tau\leq\beta$ and the fermionic Matsubara frequency $\omega_{n}=(2n+1)\pi/\beta$
with $n=0,\pm1,\pm2,\cdots$.

The local properties of the Hubbard model on the Bethe lattice can be obtained via a single-site
impurity problem supplemented by the following self-consistent relation
\cite{Caffarel1994,Laloux1994}:
\begin{equation}\label{self-consistency}
\mathcal{G}_{0\sigma}^{-1}(i\omega_{n})=i\omega_{n}+\mu-t^{2}\mathcal{G}_{\sigma}(i\omega_{n}),
\end{equation}	
where $\mathcal{G}_{0\sigma}$ is the bare Green's function. The self-consistent relation ensures that
the on-site (local) component of the Green's function
[$\mathcal{G}_{ii}(i\omega_{n})=\frac{1}{N}\sum_{k}\mathcal{G}({\bf k},i\omega_{n})$] coincides
with the Green's function $\mathcal{G}_{\sigma}(i\omega_{n}$) calculated from the effective action.

The Green's function $\mathcal{G}_{imp}(i\omega_{n})$ of the impurity model (\ref{anderson}) is
then calculated by the Lanczos method\cite{Dagotto1994,Niu2019,Amaricci2022}, which can be expressed
explicitly as \cite{Georges1996,Caffarel1994,Capone2007}
\begin{equation}
\mathcal{G}_{imp}(i\omega_{n})=\mathcal{G}_{\sigma}^{+}(i\omega_{n})
+\mathcal{G}_{\sigma}^{-}(i\omega_{n}),
\end{equation}
with $\mathcal{G}_{\sigma}^{+}(i\omega_{n})$ and $\mathcal{G}_{\sigma}^{-}(i\omega_{n})$ that are given by
\begin{eqnarray}\label{lanczos}		
\mathcal{G}_{\sigma}^{+}(i\omega_{n})=&&\frac{\langle\phi_{0}
|d_{\sigma}d_{\sigma}^{\dag}|\phi_{0}\rangle}{i\omega_{n}-a_{0}^{(+)}
-\frac{b_{1}^{(+)2}}{i\omega_{n}-a_{1}^{(+)}
-\frac{b_{2}^{(+)2}}{i\omega_{n}-a_{2}^{(+)}-\cdots}}},     \\	
\mathcal{G}_{\sigma}^{-}(i\omega_{n})=&&\frac{\langle\phi_{0}
|d_{\sigma}^{\dag}d_{\sigma}|\phi_{0}\rangle}{i\omega_{n}+a_{0}^{(-)}
-\frac{b_{1}^{(-)2}}{i\omega_{n}+a_{1}^{(-)}
-\frac{b_{2}^{(-)2}}{i\omega_{n}+a_{2}^{(-)}-\cdots}}}, ~~~~~
\end{eqnarray}
where $a_{n}$ ($b_{n}$) is the $n$th sub-diagonal element of the tridiagonalized Hamiltonian
obtained by the Lanczos method and $|\phi_{0}\rangle$ is the ground state of the
Hamiltonian (\ref{anderson}). In our calculations, we choose $n_{b}=7$ and $\beta=1024$ to assure the accuracy of the self-consistency calculations, especially in the low-energy region.
It is worth noting that $\beta$ plays a role of frequency cutoff\cite{Caffarel1994},
and hence $1/\beta$ can be regarded as a fictitious temperature.
In this work, we restrict our calculations to the zero-temperature conditions.

The Green's function behaves differently depending on whether the eigenstates are localized or extended
\cite{Economou2006}, which helps us to obtain the interaction effect on the phase transitions.
For an interaction-driven Mott transition, the ground state of the metallic phase is gapless,
while the Mott insulating ground state has a gap.
As discussed in Refs.~[\cite{Gu2009,Chen2010,Wen2017,Chen2011}],
there is a short-range entanglement in gapped quantum states, which corresponds to a symmetry protected topological (SPT) order\cite{Wen2017}.
We extend the classification method of SPT phases in higher dimensions to label-gapped
quantum phases based on the four occupation states of electrons on an impurity site.
Additionally, the fidelity per site method\cite{Zhou2008,Gu2010} is in accord with the DMFT idea of mapping a lattice model onto an effective single-site impurity model\cite{Georges1996}.
It has been demonstrated that the fidelity per site method can help us understand how quantum phase
transitions are influenced by quantum fluctuations\cite{Zhou2008,Gu2010}.
Considering the scenario of the SPT\cite{Chen2010,Gu2009,Wen2017,Chen2011,Chen2019,Wen2019}
and the sensitive feature of fidelity in detecting quantum fluctuation\cite{Gu2010},
we introduce the local quantum state fidelity\cite{Rams2011} of single impurity site as
\begin{equation}\label{mq}
L_{o}=-\frac{1}{\beta}\sum_{n=-\infty}^{\infty}e^{i\omega_{n}0^{+}}
\mathcal{G}_{imp}(i\omega_{n})\langle \Phi_{imp}^{o}|\hat{P}|\Phi_{imp}^{o\prime}\rangle,
\end{equation}
with
\begin{equation}
|\Phi_{imp}^{o}\rangle=\sum_{s=1}^{4}p_{s}|p_{s}\rangle=p_{1}|0\rangle+p_{2}|\uparrow\rangle+p_{3}|\downarrow\rangle+p_{4}|\uparrow\downarrow\rangle.\nonumber
\end{equation}
Here
$\hat{P}$ is the net spin projection operator for impurity site with $\langle0|\hat{P}|0\rangle=\langle\uparrow\downarrow|\hat{P}|\uparrow\downarrow\rangle=0$, $\langle\uparrow|\hat{P}|\uparrow\rangle=1$, and $\langle\downarrow|\hat{P}|\downarrow\rangle=-1$.
$|\Phi_{imp}^{o}\rangle$ ($|\Phi_{imp}^{o^{\prime}}\rangle$) represents the ground state wave function
of the single impurity site with an interaction strength of $U$ ($U+0^{+}$).
The factor $e^{i\omega_{n}0^{+}}$ is introduced to ensure the convergence of the summations.

In the metallic phase, the average spin of the DMFT impurity $\langle\sigma\rangle_{t}$
is zero\cite{Feng2013} because of the high symmetry of the spin at the impurity site obtained by Landau's theory.
The probabilities of doublons and holons\cite{Yuta2018} occurrence are equal, i.e., $p^{2}_{1}=p^{2}_{4}$, and the probabilities of spin-up and spin-down states have $p^{2}_{2}=p^{2}_{3}$, and therefore the LSQF keeps zero ($L_{o}$=0).
However,
the insulating ground state of the DMFT impurity model is double-degenerate with singly occupied states of opposite spin ($|\uparrow\rangle$ or $|\downarrow\rangle$), and thus $L_{o}$ has two solutions as $L_{o}=\pm C$, where C is a finite positive constant.
Because both the positive and negative signs of $L_o$ indicate the same insulating phase, we only show the absolute value of $L_o$ in the figures.
As a result, a sudden rise of the LQSF at the critical interaction $U_c$ will be found (note that $L_{o}=\pm C$ does not mean that the system is antiferromagnetic or ferromagnetic).
Specifically, the local moment of the impurity site is zero due to the double degeneration of the ground state.
Therefore, the Mott MIT can be depicted by the LQSF.
It is worth noticing that the behavior of $L_{o}$ is the same as the topological invariant found in
Ref.~[\cite{Sen2020}].

\begin{figure}[h!]
\centering
\includegraphics[scale=0.5]{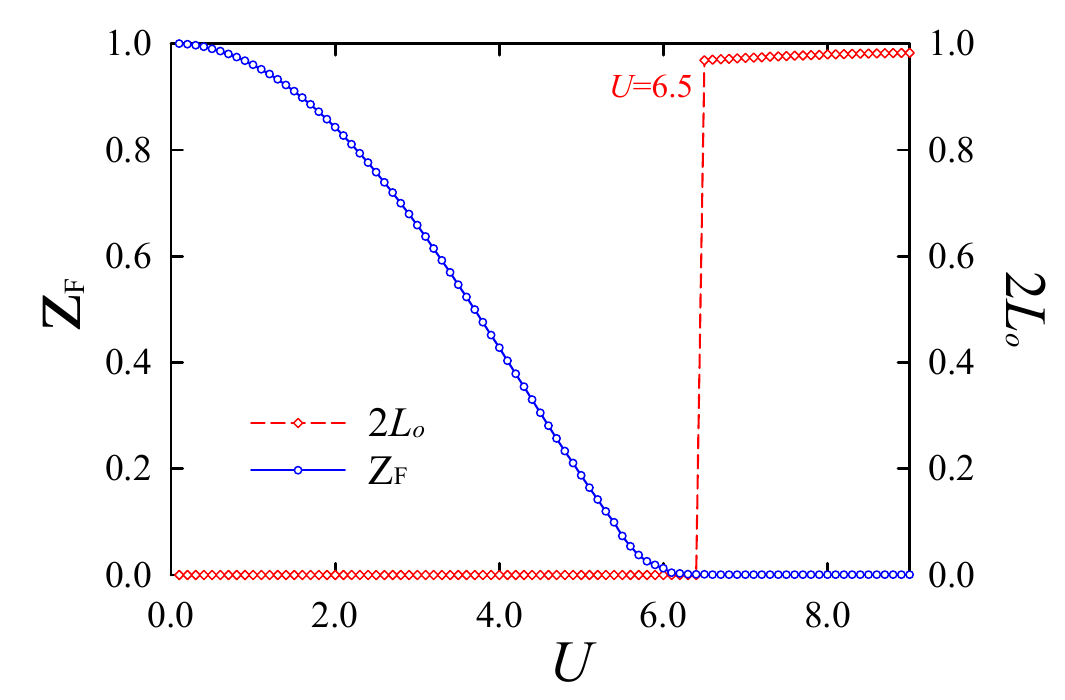}
\caption{(Color online) The local quantum state fidelity $L_{o}$ (red dotted line) as a function of interaction $U$.
The Mott metal-insulator transition occurs at a critical value of $U_c=6.5$. For comparison, the evolution of the quasiparticle coherent weight $Z_{\rm F}$ with $U$ (blue solid line) is also presented.
In the numerical calculations we chose $n_{b}=7$ and $\beta=1024$.
 \label{fig1}}
\end{figure}

To evaluate the frequency summation over the Matsubara Green's function, we need to further simplify the above formula by considering the interacting Matsubara Green's function at the poles, which holds
\begin{eqnarray}\label{p_and_n}
[\mathcal{G}_{\sigma}(i\omega_{n})]&=&[\mathcal{G}_{\sigma}(-i\omega_{n})]^{*},\nonumber \\
\omega_{n}&=&\frac{(2n+1)\pi}{\beta},\quad n=0,1,2,\cdots.
\end{eqnarray}
With the help of the above equation (\ref{p_and_n}), the LQSF of the impurity site
can be rewritten explicitly as, $L_{o}=-\frac{2}{\beta}\rm{Re}\sum_{n=0}^{\infty}e^{\it{i}\omega_{n}0^{+}}$
$\mathcal{G}_{imp}({\it i}\omega_{n})\langle{\Phi_{imp}^{o}|\hat{P}|\Phi_{imp}^{o\prime}}\rangle$,
indicating that $L_{o}$ can be directly obtained by a summation of the positive frequencies in the effective on-site problem.

\section{Results}
We define the quasiparticle coherent weight $Z_{\rm F}$ as \cite{Mahan2000,Perroni2007}
\begin{eqnarray}
\frac{1} {Z_{\rm F}}=1-\frac{\partial}{\partial\omega}\rm{Re}\Sigma_{\sigma}(\omega)|_{\omega=0}
\approx 1-\frac{\rm{Im}\Sigma_{\sigma}({\it i}\omega_{0})}{\omega_{0}}, ~~~~~
\end{eqnarray}
where the local self-energy $\Sigma_{\sigma}(i\omega_{n})$ is obtained from the local Green's
function in Eq. (\ref{self-consistency}). In the following discussions, we study the Mott MIT
of the Hubbard model at half-filling in terms of the evolution of the LQSF $L_{o}$
at the impurity site with the on-site Coulomb repulsive interaction $U$.
In Fig.~\ref{fig1} we plot $2L_{o}$ as a function of interaction $U$, where the red dashed-line indicates the position of the
critical point of the Mott MIT. For a better comparison, the evolution of the quasiparticle
coherent weight $Z_{\rm F}$ (blue solid-line) with $U$ is also presented in Fig.~\ref{fig1} .
$Z_{\rm F}$ usually decreases with increasing $U$ and keeps very close to zero when approaching
the critical interaction $U_c$ of the Mott MIT, near which the systematic errors of $Z_{\rm F}$ increases
significantly, leading to difficulties in the quantitative determination of the critical point of MIT.
More crucially, within the framework of DMFT, two metallic results with different slope $dZ_F/dU$
are found to coexist in a finite range of interaction strengths \cite{Chatzieleftheriou2022,Ono2003},
and thus a comparing of the respective energies with the energy of the insulator is suggested\cite{Ono2003}.
\begin{figure}[h!]
\centering
\includegraphics[scale=0.53]{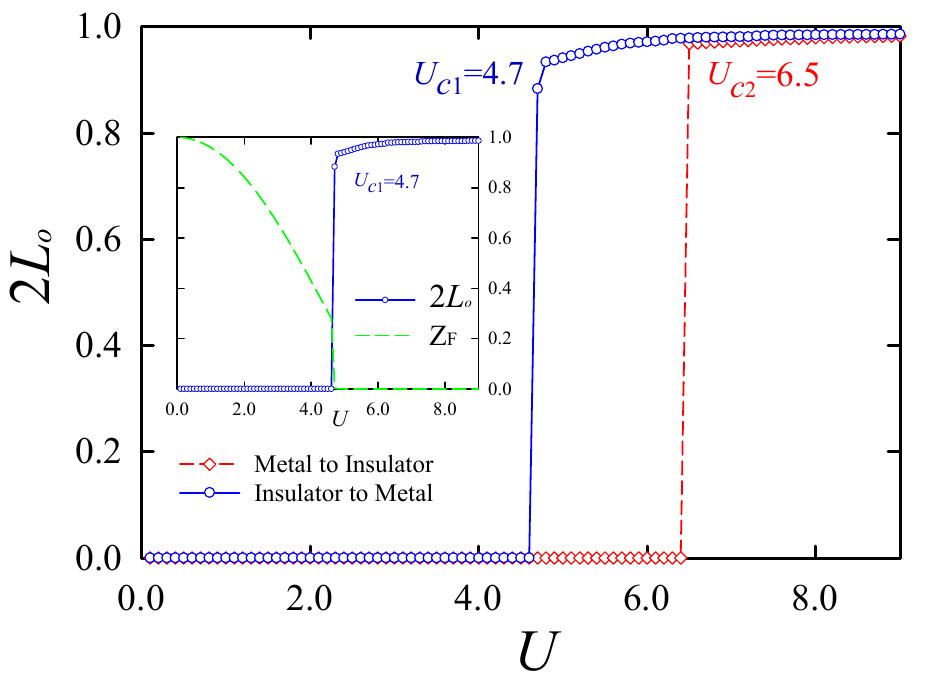}
\caption{(Color online) The local quantum state fidelity $L_{o}$ as a function of interaction $U$.
  The blue line indicates the critical point at $U_{c_{1}}=4.7$, while the red line denotes the
  critical point at $U_{c_{2}}=6.5$. Within the region between the two critical points of $U_{c_{1}}$
  and $U_{c_{2}}$, the insulating solution (blue solid line) and metallic solution (red dotted line)
  coexist. Inset: The comparison of the local quantum state fidelity (blue solid line) and the quasiparticle coherent
  weight (green dotted line) for the insulating-phase solution. \label{fig2}}
\end{figure}
Consequently, it becomes quite difficult and inconvenient to numerically determine the actual critical interaction $U_c$ by the quasiparticle coherent weight.
In a striking contrast to the complex of $Z_{\rm F}$ that has two metallic solutions in the coexistence
region of interaction $U$, the LQSF $L_{o}$ keeps equal to zero for both the metastalbe and
stable metallic solutions when $U<U_c$, as shown in Fig.~\ref{fig1}.
However, at the critical point $U_c$=6.5 of MIT, the LQSF jumps abruptly from $L_{o}=0$ in
the metallic phase to $2L_{o}\approx 1.0$ in the insulating phase.
Our results indicate clearly that the
LQSF $L_{o}$ at the impurity site is very sensitive to the existence of the resonant peak
at the Fermi level, which therefore is a more proper physical quantity to quantitatively depict the critical point of the Mott MIT.

As to the two classes of solutions, (i) the solution from the metallic phase towards the critical interaction of MIT
(the metallic-phase solution $U_{c2}$ ) and (ii) the solution from the insulating phase
towards the critical point of MIT (the insulating-phase solution $U_{c1}$ ), the LQSF $L_{o}$ as a more
proper physical quantity of MIT can give a natural explanation of the difference between the
metallic and insulating solutions.
In this case, we have made a series of calculations for $2L_{o}$, and the results of the metallic solution of $2L_{o}$ (red dotted-line) and the insulating solution (blue solid-line) are plotted in Fig.~\ref{fig2}, where $U_{c1}=4.7$ and $U_{c2}=6.5$ are the critical points of the insulator-to-metal transition and the metal-to-insulator transition, respectively.
\begin{figure}[h!]
\centering
\includegraphics[scale=0.53]{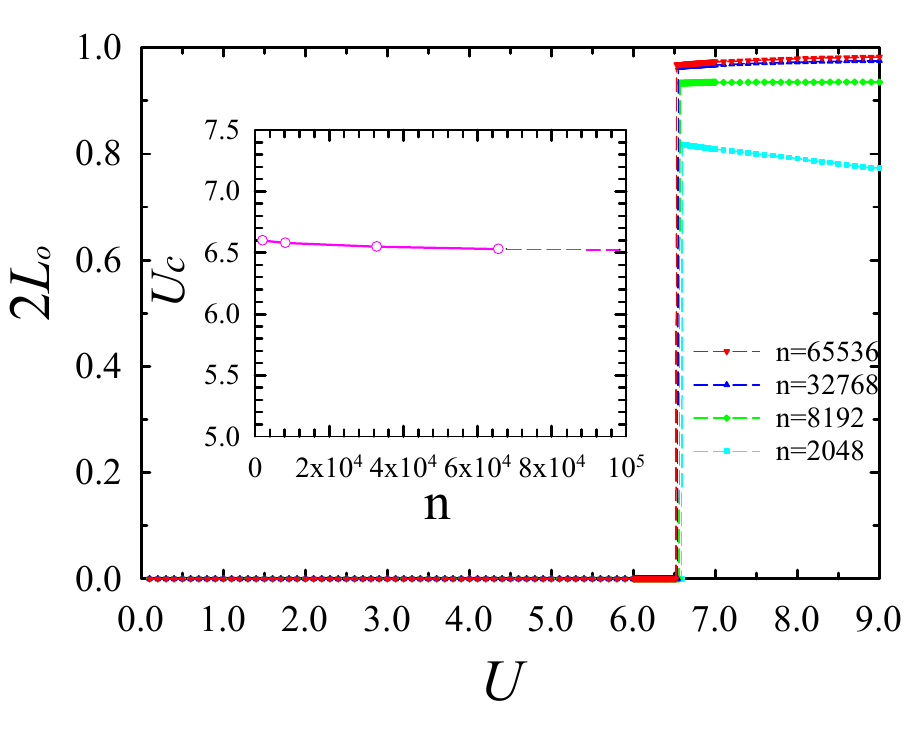}
\caption{(Color online)
The local quantum state fidelity $L_{o}$ as a function of the
interaction $U$ for various cutoff values $n$ in the series summation.
The evolution of $U_c$ with $n$ (circles) is presented (inset) along with the fitting line (solid line) and its extension
(dashed line). \label{fig3}}
\end{figure}
Our findings of the critical interactions are in agreement with the DMFT results
from the numerical renormalization group solver\cite{Bulla1999} $U_{c1}=5.0$ and $U_{c2}=5.88$,
and the dynamical density renormalization group method\cite{Karski2005} $U_{c1}=4.76$ and $U_{c2}=6.14$.
The results in Fig.~\ref{fig2} show that apart from a metallic phase in the weak interaction
region ($U<U_{c1}$) and an insulating phase in the strong interaction region ($U>U_{c2}$), there
is an intermediate interaction region ($U_{c1}<U<U_{c2}$), where the metallic solution coexists
with the insulating solution. Within this intermediate interaction region, $2L_{o}$ as a
function of $U$ exhibits a hysteretic behavior since both the metallic and insulating solutions
are found to be attractive points of a particle-hole symmetry system \cite{Loon2020}.
The present results in Fig.~\ref{fig1} and Fig.~\ref{fig2} therefore show that $L_{o}$ is a more proper physical parameter to give a quantitative description of the Mott MIT in a strong correlation system.

Although the summation of Matsubara frequency is from zero to infinity, the actual
calculation is performed numerically with the infinitude of Matsubara frequency $n=0,1,2,...,
\infty\rightarrow n=0,1,2,...,n_{max}$ replaced by a finite $n_{max}$. In this case, we have
made a series of calculations for $2L_{o}$ as a function of $U$ at different cutoff $n_{max}$,
and the results are plotted in Fig.~\ref{fig3}, where the critical points at $n_{max}=2048$,
$n_{max}=8192$, $n_{max}=32768$, and $n_{max}=65536$ are very close to each other, indicating
that for the large enough $n_{max}$, the error bars are small enough. In particular, $U_c$ can
be extrapolated as $U_c=6.5$ in the case of $n_{max}=\infty$.

\begin{figure}[h!]
\centering
\includegraphics[scale=0.53]{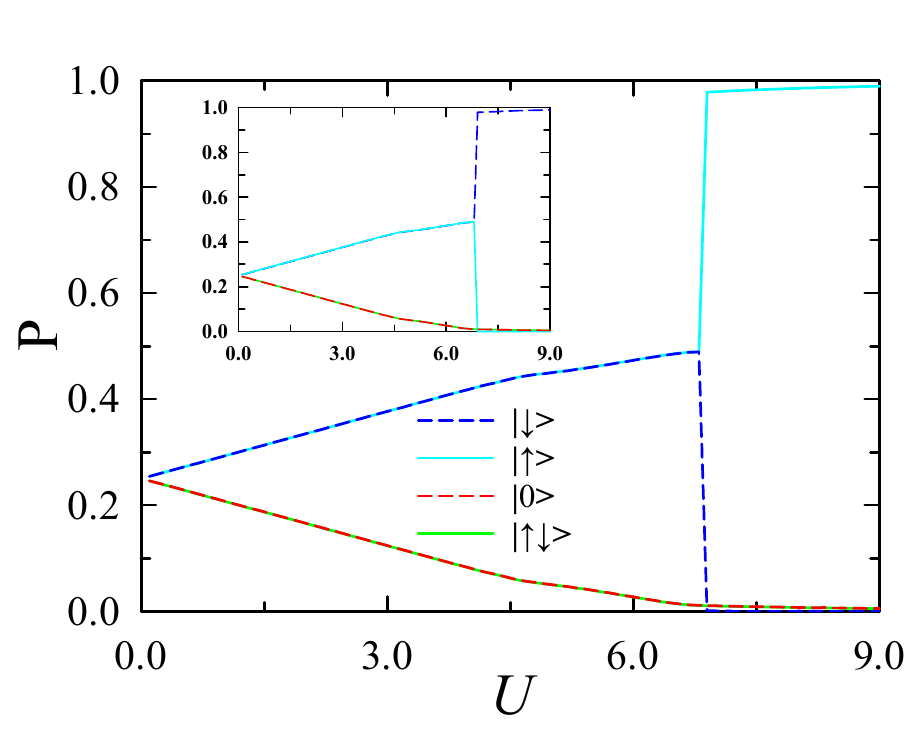}
\caption{(Color online) The probabilities $p^{2}$ of the zero occupied state $|0\rangle$ (red dashed
line), spin-up occupied state $|\uparrow\rangle$ (cyan solid line), spin-down occupied state
$|\downarrow\rangle$ (blue dashed line), and double occupied state $|\uparrow\downarrow\rangle$
(green solid line) in the impurity ground state as a function of the interaction $U$.
Inset: another solution with opposite $p^{2}$ of the spin-up and spin-down occupied states
when $U>U_c$, demonstrating the double-degeneration of the ground state within the insulating phase.
\label{fig4}}
\end{figure}

The Hilbert space of each site in the Hubbard model (\ref{hubbard-model}) consists of four states,
$|0\rangle$, $|\uparrow \rangle$, $|\downarrow \rangle$, $|\uparrow\downarrow\rangle$, corresponding
to the zero, spin-up, spin-down, and double-electron-occupied states, respectively.
The probabilities of the zero, spin-up, spin-down, and double-occupied states at the single impurity site of the metallic solution
are plotted in Fig.~\ref{fig4}, which shows clearly that the probabilities of the zero and double-occupied states are equal and decrease simultaneously in the metallic phase.
However, the probability of the spin-up singly occupied state increases with the increase of the on-site Coulomb interaction $U$ and jumps to
$p^{2}\approx 1$ at the critical point $U_{c2}=6.5$.
The same feature for the probability of the spin-down singly occupied state
is found in the insulating phase due to the degeneration(see Fig.~\ref{fig4} inset).
Concomitantly, the LQSF is equal to zero when the probability of the spin-up singly occupied state is equal to that of the spin-down
singly occupied state in the metallic phase.
However, in a striking contrast to the case in the metallic phase, the feature of the probability of
the spin-up singly occupied state is quite different from that of the probability of the spin-down
singly occupied state in the insulating phase, and a jump of the LQSF is found at the
critical point due to the presence of two degenerate solutions with opposite spin occupancies.
The above results correspond to the theoretical prediction of Eq (\ref{mq}).
This is why the LQSF in Eq (\ref{mq}) is a more proper
physical quantity to give a quantitative depiction of two distinct forms of the critical points in
MIT of a strongly correlated system.

\section{Conclusions}
Based on the one-band Hubbard model, we have studied the Mott MIT in a strongly
correlated system by using the combined approach of the dynamical-mean field theory and Lanczos technique.
Our results clearly demonstrate that the local quantum state fidelity serves as a proper physical quantity
for depicting the Mott metal-insulator transition in a strongly correlated system.
It allows for quantitatively determining of the critical points and provides a
consistent description of two distinct forms of the critical points.
The local quantum state fidelity can be also used to discuss the novel
physics in orbital-selective Mott insulators \cite{Kotliar2006} and superconductors
\cite{Le2006,Neto2009}. In particular, it may be applied to explain the hysteresis observed
experimentally in Mott-field effect transistors \cite{Kim2004}. These related
works are currently under research.

\section*{Acknowledgements}

The authors would like to thank Gabriele Bellomia for fruitful discussions. YN is also grateful
to Louk Rademaker and Haiming Dong for helpful discussions.
Project supported by the Scientific Research Foundation for Youth Academic Talent of Inner Mongolia
University (Grant No.10000-23112101/010) and the Fundamental Research Funds for the Central Universities of China (Grant No. JN200208).
YS is supported by the National Natural Science Foundation of China (Grant No. 11474023).
SF is supported by the National Key Research and Development Program of China (Grant No. 2021YFA1401803) and the National Natural Science Foundation of China (Grant Nos. 11974051 and 11734002).

\bibliography{apssamp}

\end{document}